\documentclass[twocolumn]{revtex4}

\usepackage{bm}

\usepackage{graphicx}

\usepackage{subfigure}
\usepackage{setspace}

\newcommand{\simleq}{\; \raisebox{-0.4ex}{\tiny$\stackrel
{{\textstyle<}}{\sim}$}\;}
\newcommand{\simgeq}{\; \raisebox{-0.4ex}{\tiny$\stackrel
{{\textstyle>}}{\sim}$}\;}

\begin{document}

%\setstretch{2}

\title{Short-Time Enhanced Decay of Coherent Excitations in Bose-Einstein Condensates}

%% Notice placement of commas and superscripts and use of &
%% in the author list

\author{Nir Bar-Gill}
%\affiliation{Dept. of Physics of Complex Systems, Weizmann Institute of Science, Rehovot 76100, Israel}
\author{Eitan E. Rowen}
%\affiliation{Dept. of Physics of Complex Systems, Weizmann Institute of Science, Rehovot 76100, Israel}
\author{Gershon Kurizki}
%\affiliation{Dept. of Physical Chemistry, Weizmann Institute of Science, Rehovot 76100, Israel}
\author{Nir Davidson}
%\affiliation{Dept. of Physics of Complex Systems, Weizmann Institute of Science, Rehovot 76100, Israel}

\affiliation{Weizmann Institute of Science, Rehovot 76100, Israel}

\begin{abstract}
We study, both experimentally and theoretically, short-time modifications of the decay of excitations in a Bose-Einstein Condensate (BEC) embedded in an optical lattice. Strong enhancement of the decay is observed compared to the Golden Rule results. This enhancement of decay increases with the lattice depth. It indicates that the description of decay modifications of few-body quantum systems also holds for decay of many-body excitations of a BEC.
\end{abstract}

\maketitle

%\spacing{2}

Relaxation and decoherence of excited quantum states, caused by their coupling to the environment, is a ubiquitous phenomenon. Yet it still poses a fundamental
question: can we describe its effects on complex, {\em many-body} quantum states, as in simple quantum systems? In few-body systems, control by fast modulation or short-time duration may result in either relaxation slowdown (quantum Zeno effect - QZE) or its speedup (anti-Zeno effect - AZE) \cite{fonda,kurizki,raizen,pascazio} compared to the Golden Rule (GR) result \cite{sakurai}. Here we
experimentally and theoretically study these concepts for the decay of excitations in a Bose-Einstein Condensate (BEC). To this end, we apply the control methods of
short-time duration and reservoir engineering \cite{myatt,yablonovitch} to momentum excitations of a BEC in a deep one-dimensional (1D) optical lattice. 
%We find enhancement of the decoherence as a function of the lattice depth, thereby demonstrating that the universal approach developed for {\em simple} systems \cite{kurizki} is also valid in a many-body, decaying quantum system.

We first use two-photon Bragg spectroscopy \cite{bragg,ozeri} to characterize the excitation spectrum at different lattice  depths. We then measure the decay rate of these excitations, a process describable as the
Beliaev damping of Bloch-Bogoliubov states \cite{kramer,beliaev}. We find the decay rate to increase with lattice depth. This increase is in contradiction with the long-time GR theory \cite{griffin}. We show that the non-exponential (AZE) regime predicted by the universal theory of short-time decay \cite{kurizki}, and hitherto observed for {\em single atoms} in optical lattices \cite{raizen}, is in good agreement with the experimental results for this {\em many-body} process.

\begin{figure}
%\subfigure[]{
%\includegraphics[width=0.48\linewidth,height=0.35\linewidth]{new2dsetup6}}
%\subfigure[]{
%\includegraphics[width=0.48\linewidth,height=0.35\linewidth]{lattice_calib_bragg_noint_qy118_with_pulse4}}
\subfigure[]{
\includegraphics[width=0.48\linewidth,height=0.35\linewidth]{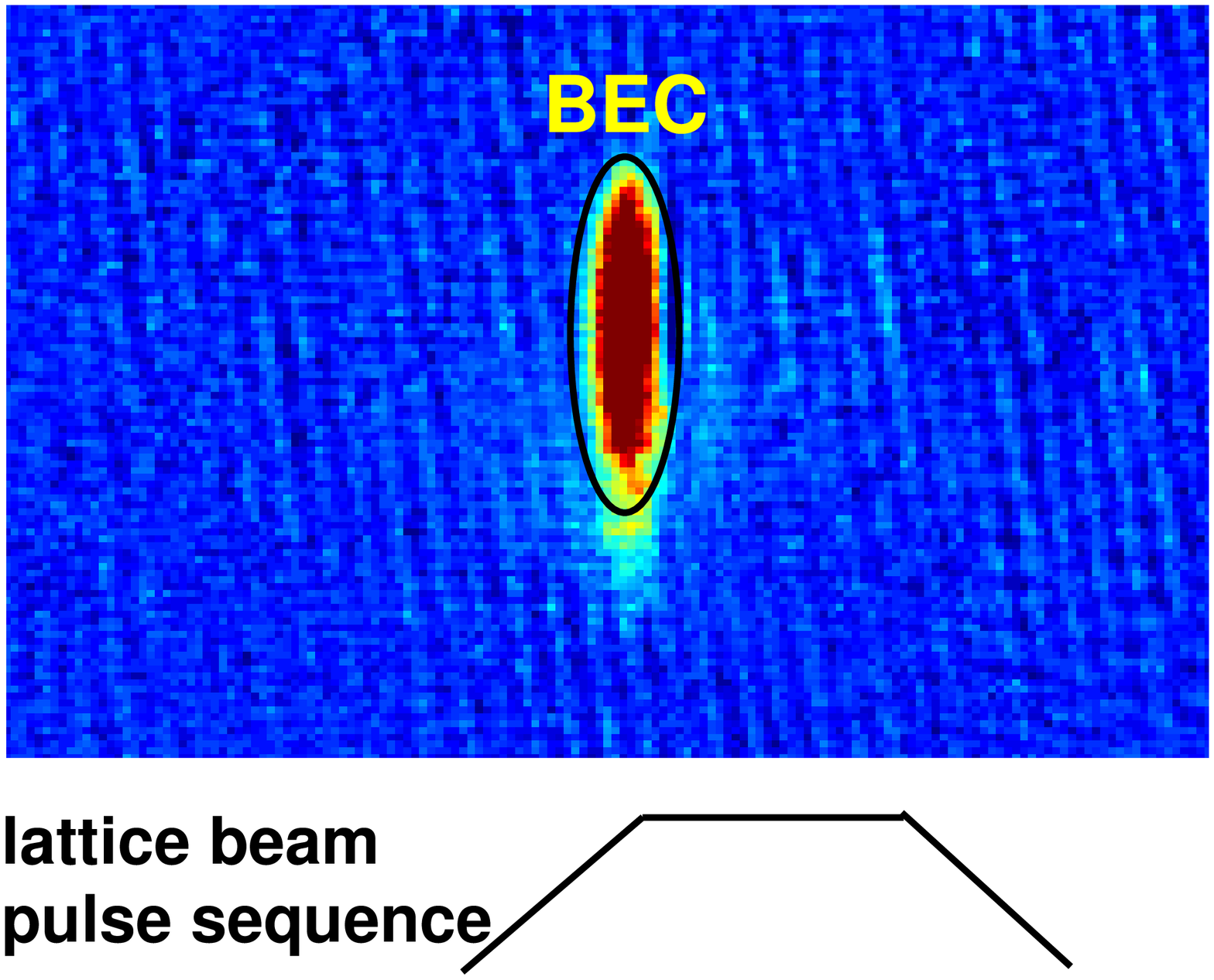}}
\subfigure[]{
\includegraphics[width=0.48\linewidth,height=0.35\linewidth]{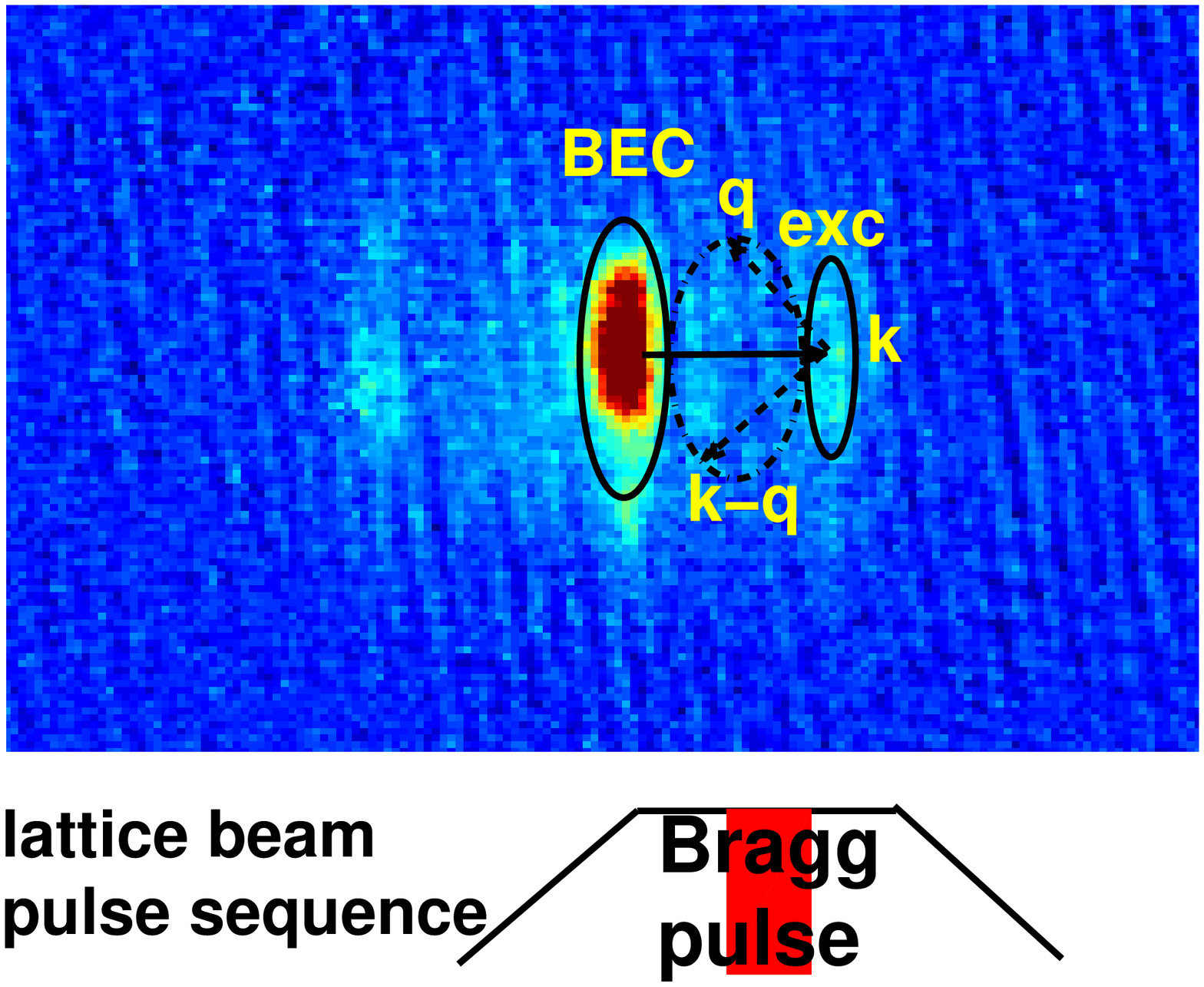}}%{bragg_in_lattice_1_2V_20mV_new_arrows_pulse5}}
\caption{%(a) Schematic view of the experimental setup.
%absorption imaging is used to take a time of flight image of their momentum distribution.
%(b) A time-of-flight image after $38$ ms of a sequence in which the lattice beams are pulsed on and off, diffracting the atoms to momentum $\pm 2 k_L$.
(a) A momentum-space absorption image ($38$ ms time-of-flight) of the adiabatic switching on and off of an optical lattice of depth $5 E_R$. Similar images are obtained for deep lattices as well ($V_{lat} \simeq 50 E_R$). 
(b) A $38$ ms time-of-flight image of the Bragg excited BEC with $k=0.9 k_L$, inside a lattice of $5 E_R$. 
% Its momentum distribution includes components at both $k=0.9 k_L$ and at $k=-1.1 k_L$, which are visible in the image. The image also exhibits weakly populated modes at $|q|<|k|$ due to decay, which are indicated by the $q$ and $k-q$ arrows.
} \label{fig:setup}
\end{figure}

Our experiment involves a BEC of $\sim 300,000$ $^{87}$Rb atoms in state $|F=2,m_F=2 \rangle$, held in a cigar-shaped magnetic trap, with
trapping frequencies $\omega_{\perp}=350$ Hz, $\omega_z=30$ Hz. The chemical potential is $\sim 4$ kHz, and the temperature
 is smaller than $100$ nK.
To alter or engineer the excitation spectrum of the BEC, we adiabatically load it onto a 1D optical lattice \cite{denschlag}, created by two
far-detuned ($0.5$ nm) laser beams with wavevector $k_L$, counter-propagating along the z direction (Fig. \ref{fig:setup}(a)).
% (Fig. \ref{fig:setup}(a)). 
Our adiabatic ramp-up time of $250 \mu$s is much longer than the timescale of the adiabaticity condition $dV_{lat}/dt << 16 E_R^2/\hbar$ \cite{denschlag}, which is $\sim 125 \mu$s for $V_{lat} \simeq 50 E_R$. This ensures adiabaticity with respect to higher band excitations, while the many-body dynamics and trap dynamics are nearly frozen-out. These conditions allow us to study the dynamics of momentum excitations of a condensate adiabatically loaded onto a 1D lattice. In the Supporting Information we quantitatively analyze the non-adiabatic effects as a function of lattice depth, following which we limit our experiment to lattices of $V_{lat} \simleq 50E_R$.
%, in the framework of Bogoliubov theory.
%Adiabatic loading of the BEC into the
%ground-state of the lattice potential is verified by adiabatically turning the lattice on and then off, and recovering a pure BEC in the
%time-of-flight (momentum space) images (Fig. \ref{fig:setup}(a)). We calibrate the lattice depth by suddenly pulsing the lattice beams on and off for $20 \mu$s and measuring the fraction of atoms diffracted into the $\pm 2k_L$ modes.% (Fig. \ref{fig:setup}(b)).
Once the BEC is thus loaded into the lattice ground state, we coherently
excite it close to the Brillouin-zone edge by a pair of Bragg beams (detuned from resonance by $0.2$ nm). This two-photon Bragg spectroscopy \cite{bragg,ozeri} allows us to excite and probe the BEC at a given momentum $k=0.9 k_L$ (along the z direction) and energy $E_k$.
%We then monitor, by momentum-space absorption images, the number of atoms remaining coherently excited as a function of time following the
%excitation.
The time-of-flight image after the Bragg excitation (Figure \ref{fig:setup}(b)) shows the condensate, the excitation and the decay products (occupied modes with $|q|<|k|$), created through collisions between the BEC and the excitation. 
%Note that since the excitation is close to the Brillouin-zone edge, during the ramp-down of the lattice, which is not adiabatic with respect to the excitation, a momentum component at $k=-1.1 k_L$ appears. This is taken into account in the analysis below.
%Figure \ref{fig:setup}(d)depicts a time-of-flight image after such a Bragg excitation. The condensate, the excitation and the decay products, created through collisions between the BEC and the excitation, are all evident in the image (the decay products appear as occupied modes with $|q|<|k|$).

The perturbative excitations of the BEC are collective (many-body) quasi-particles
%which obey the theoretical description of Bogoliubov \cite{bogoliubov} 
in the optical-lattice Bloch-state basis.
They can be analyzed by 3D Bloch-Bogoliubov equations \cite{kramer}
%\footnote{The depletion of the condensate in our experiment is bounded by a few percent, since the number of atoms per lattice site is large (see \cite{kramer}). Hence, the Bogoliubov approach used in the paper is valid.} 
in terms of the condensate wavefunction
$\Psi$, embedded in a periodic potential in the axial direction, and assuming planar excitations in the transverse
($\bot=x,y$) directions \footnote{See Supporting Information, Sec. II, par. 2.}:
\begin{eqnarray}
\left [ - \frac{\hbar \partial_z^2}{2 m} -\frac{\hbar k_\bot^2}{2 m} + V_{lat} sin^2 (k_L z ) - \mu + \frac{2 g n \pi}{k_L} | \Psi |^2 \right] u_{jq} (z) \nonumber \\
+ \frac{g n \pi}{k_L} \Psi^2 v_{jq} (z) = \hbar \omega_j (q) u_{jq} (z), \nonumber \\
\left [ - \frac{\hbar \partial_z^2}{2 m} -\frac{\hbar k_\bot^2}{2 m} + V_{lat} sin^2 (k_L z ) - \mu + \frac{2 g n \pi}{k_L} | \Psi |^2 \right] v_{jq} (z) \nonumber \\
+ \frac{g n \pi}{k_L} \Psi^{*2} u_{jq} (z) = - \hbar \omega_j (q) v_{jq} (z). \label{bloch_bogo}
\end{eqnarray}
Here $m$ is the mass of a $^{87}$Rb atom, $V_{lat}$ is the depth of the optical lattice, $g$ is the interaction strength,
$n$ is the 3D density of the BEC, $k_\bot$ is the transverse momentum and $u_{jq}$ and $v_{jq}$ are the amplitudes of the atomic components of a
Bogoliubov excitation in Bloch band $j$ with quasi-momentum $q$. 
%In the axial ($z$) direction, due to the periodicity of the external potential, we employ Bloch theory 
%to rewrite the excitations in terms of Bloch waves as $u_{jq} (z) = e^{i q z/\hbar} \tilde{u}_{jq} (z)$,
%and recover the band structure of the spectrum. 
Solving Eqs. (\ref{bloch_bogo}) self-consistently, we find the excitation spectrum $\omega_j(q,k_\bot)$, and the amplitudes $u_{jq}(z)$ and $v_{jq}(z)$.

\begin{figure}[ht]
\subfigure[]{
\includegraphics[width=0.48\linewidth,height=0.35\linewidth]{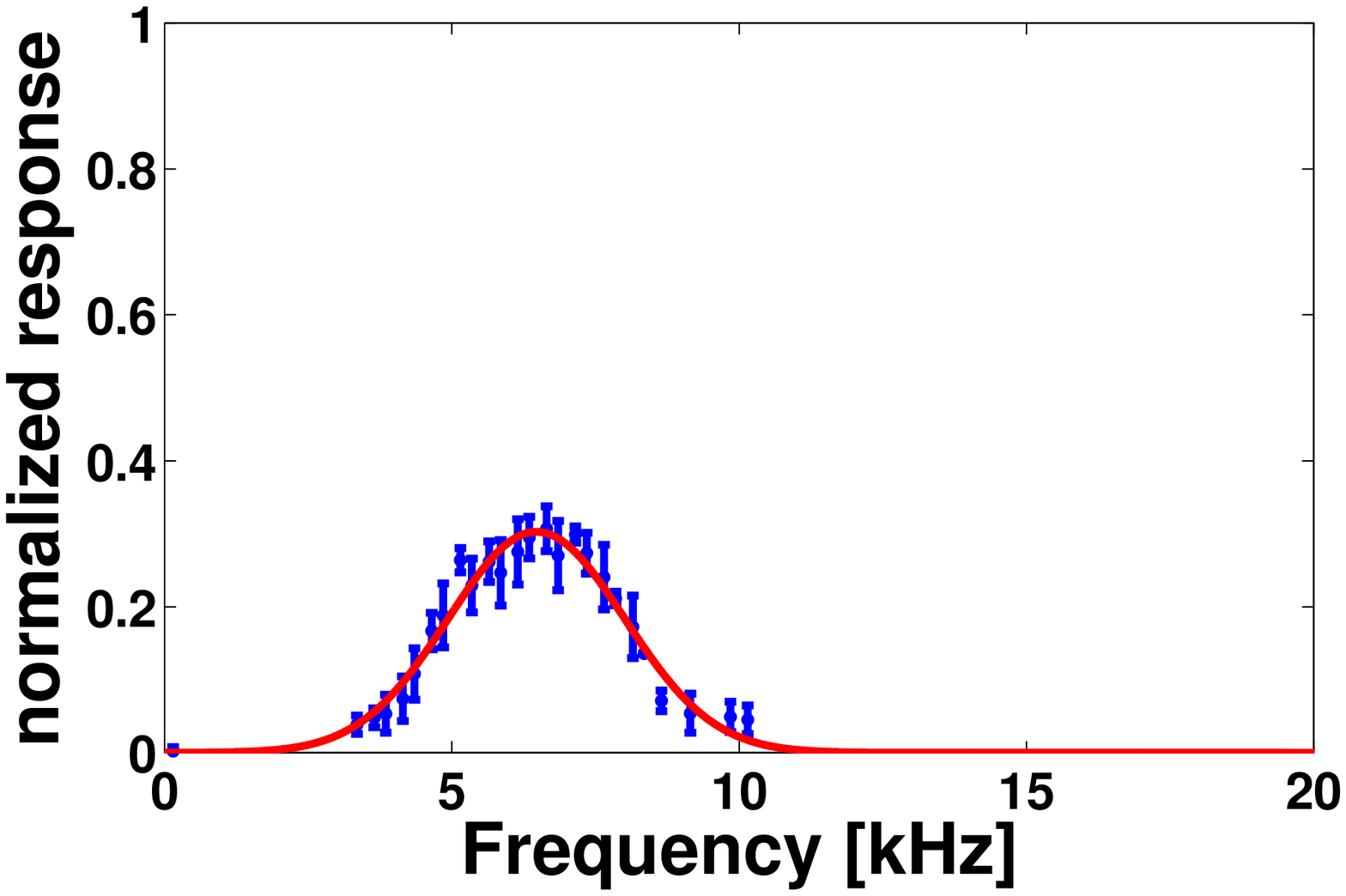}}
\subfigure[]{
\includegraphics[width=0.48\linewidth,height=0.35\linewidth]{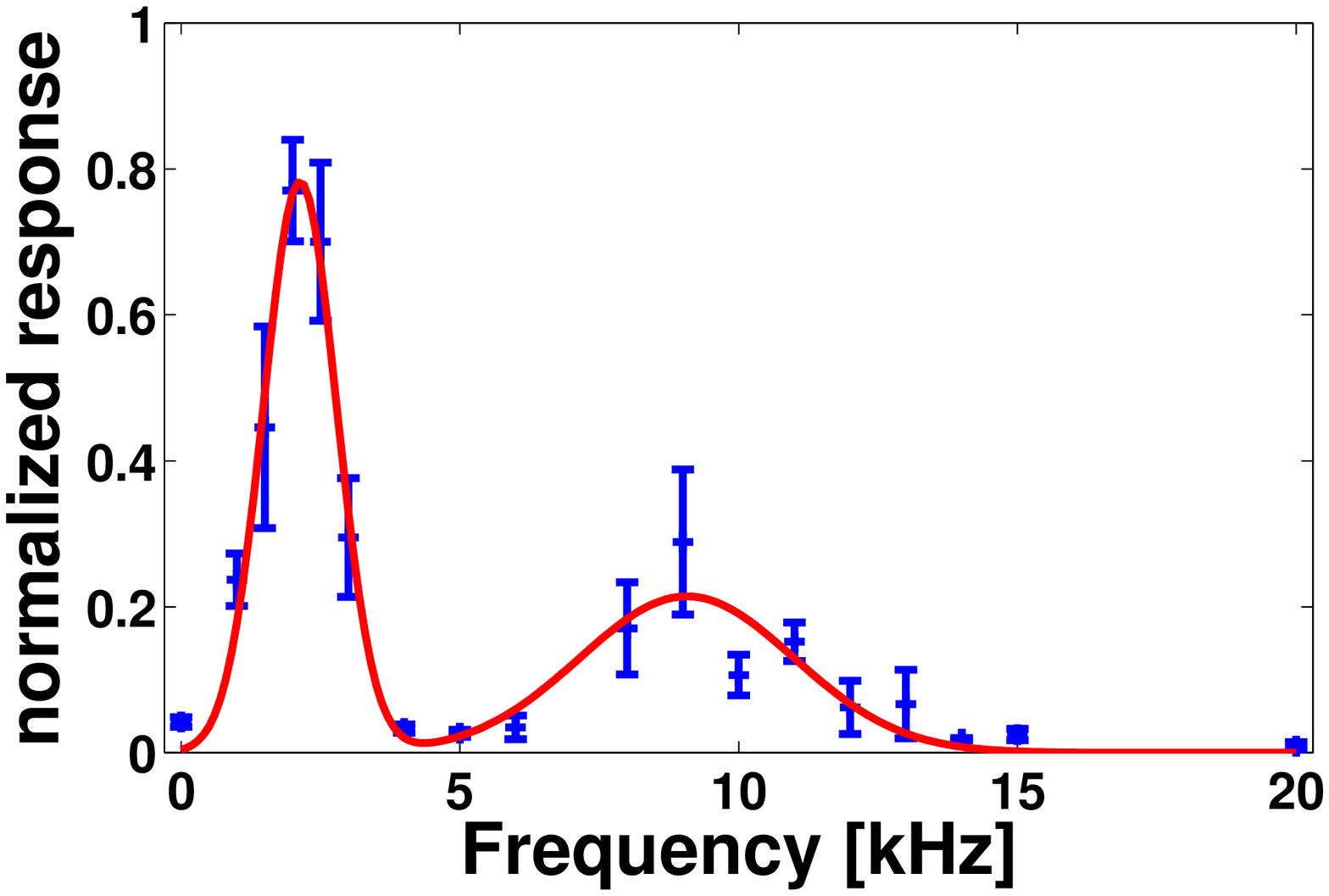}}
\caption{Excitation lineshapes measured as a function of frequency using a $2$ms Bragg pulse of $k=0.9 k_L$, without a lattice (a) and in an optical lattice of depth $V_{lat}=10 E_R$ (b). The averaged data points are shown in blue, with error-bars depicting the error of the mean. The red lines are gaussian
fits to the experimental data. 
%Without a lattice we obtain a single peak resonance lineshape centered at the expected Bogoliubov energy of the excitation. In the presence of a lattice the lineshape exhibits a double-peak structure.
} \label{fig:spect0.9kl}
\end{figure}
In Fig. \ref{fig:spect0.9kl} we plot the measured excitation lineshape for $k=0.9 k_L$ as a function of the excitation frequency $\omega$, with and without an optical lattice potential. Without the lattice, the lineshape has a single peak at the frequency of the Bogoliubov excitation (Fig. \ref{fig:spect0.9kl}(a)). In the presence of the lattice, it exhibits a double-peak structure, centered at the frequencies of the Bloch-Bogoliubov excitations of
the first and second bands, as calculated from Eq. (\ref{bloch_bogo}) (Fig. \ref{fig:spect0.9kl}(b)). For deep lattices the bands are spectrally separated, and can therefore be specifically addressed using Bragg spectroscopy. We note that these lineshapes indicate a qualitative difference in the response of the system as a function of lattice depth. This does not affect the results of this paper due to the normalization method used (see below). %\footnote{Detailed measurements of the structure factor $S(k,\omega)$ as a function of lattice depth (see \cite{kramer}) will be published in a future work.}

\begin{figure}
\includegraphics[width=0.98\linewidth,height=0.5\linewidth]{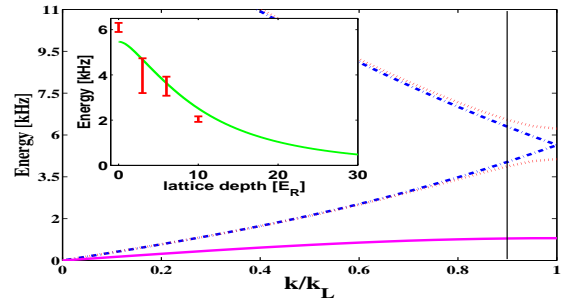}
\caption{Calculated Bloch-Bogoliubov spectrum of the first and second bands for different lattice depths ($V_{lat}=0 E_R$ - blue dashed-dotted,
$V_{lat}=1 E_R$ - red dotted, $V_{lat}=15 E_R$ - magenta solid). As the lattice depth increases, the energy band
flattens out exponentially. 
Inset: measured
(red) and calculated (green) resonant energy of the first-band $k=0.9 k_L$ quasi-momentum excitation as a function of lattice depth
(compare to the vertical line in the main figure).}
\label{fig:spect}
\end{figure}
In Fig. \ref{fig:spect} we plot the calculated Bloch-Bogoliubov excitation spectrum for lattices of different depths, and transverse momentum
$k_{\bot}=0$. For weak lattices of depth $V_{lat} \simleq E_R$ ($E_R = \hbar^2 k_L^2 /2m$ is $E_R$ the recoil energy), the lowest band spectrum resembles the standard Bogoliubov spectrum: a linear phonon branch at low momenta that becomes parabolic at high momenta \cite{bogoliubov,ozeri}. As the lattice becomes deeper, for $V_{lat} > E_R$, the energy splitting near the Brillouin zone edge curves the parabolic spectrum toward a constant value, eventually making it completely convex at $V_{lat} \simgeq 4 E_R$. Correspondingly, the width of the
lowest band decreases exponentially, according to the relation \cite{kramer} $\hbar \omega_1(k_L) \propto E_R e^{-2 \sqrt{V_{lat}/E_R}}$.

The significant narrowing of the first band as a function of lattice depth is clearly visible in Fig. \ref{fig:spect}(inset), in which we plot the calculated and measured excitation energy for $k=0.9k_L$ in the first band as a function of lattice depth. 
In the following, we concentrate on excitations in the first band, and therefore only study lattices deeper than $V_{lat} = 2 E_R$. This allows us to spectrally separate the bands using Bragg spectroscopy. 
%to ensure that we do not excite higher bands.

The coherent quasi-particle mode created by Bragg spectroscopy interacts with the BEC and decays into the continuum of other quasi-particle modes. The decay of this many-body coherent excitation into pairs of quasi-particle modes forming a continuum is governed by the Beliaev three-wave-mixing Hamiltonian \cite{ozeri}, and is known as Beliaev decay \cite{beliaev,katz_beliaev}.

\begin{figure}
\subfigure[]{
\includegraphics[width=0.48\linewidth,height=0.35\linewidth]{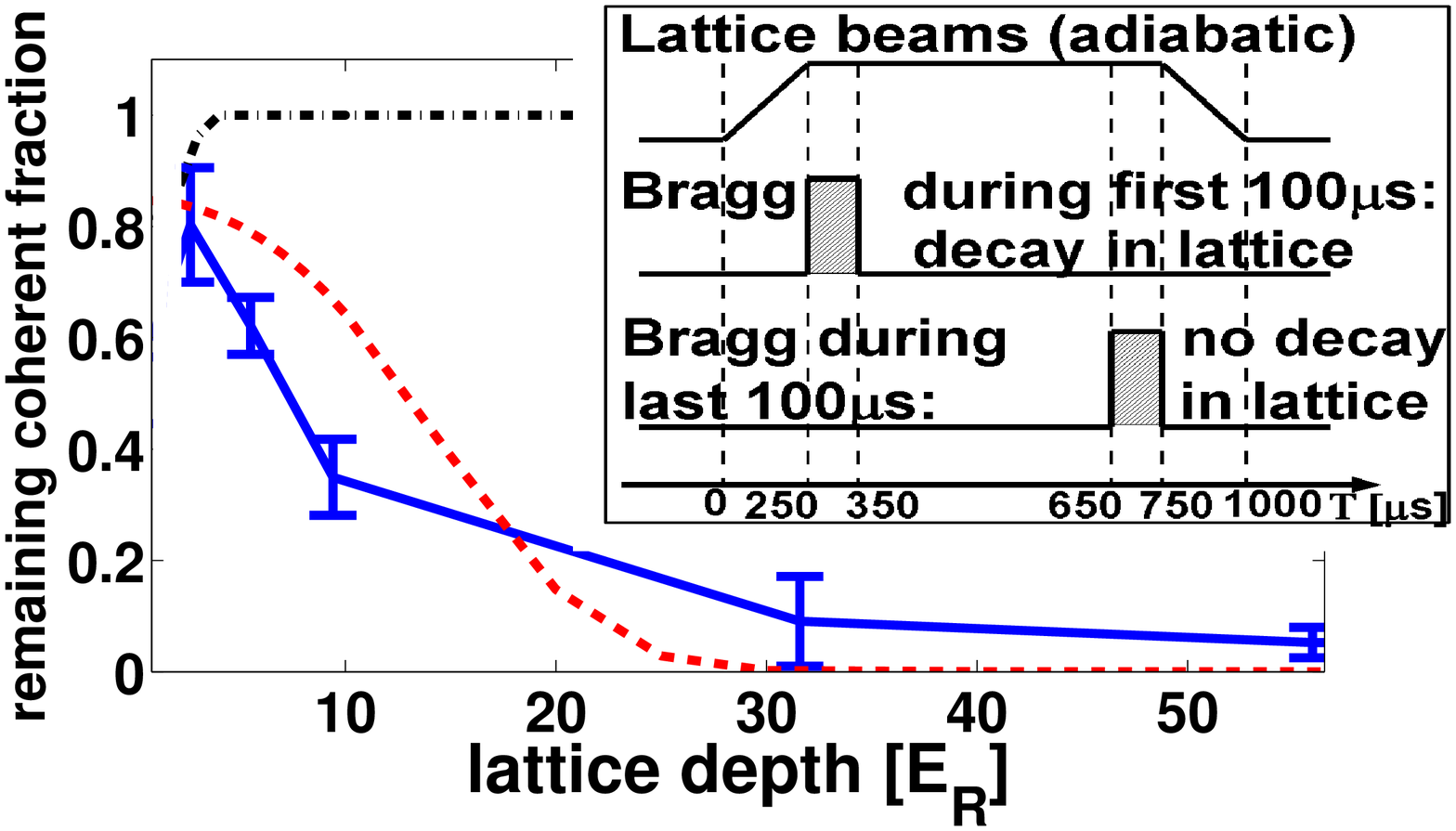}}
\subfigure[]{
\includegraphics[width=0.48\linewidth,height=0.35\linewidth]{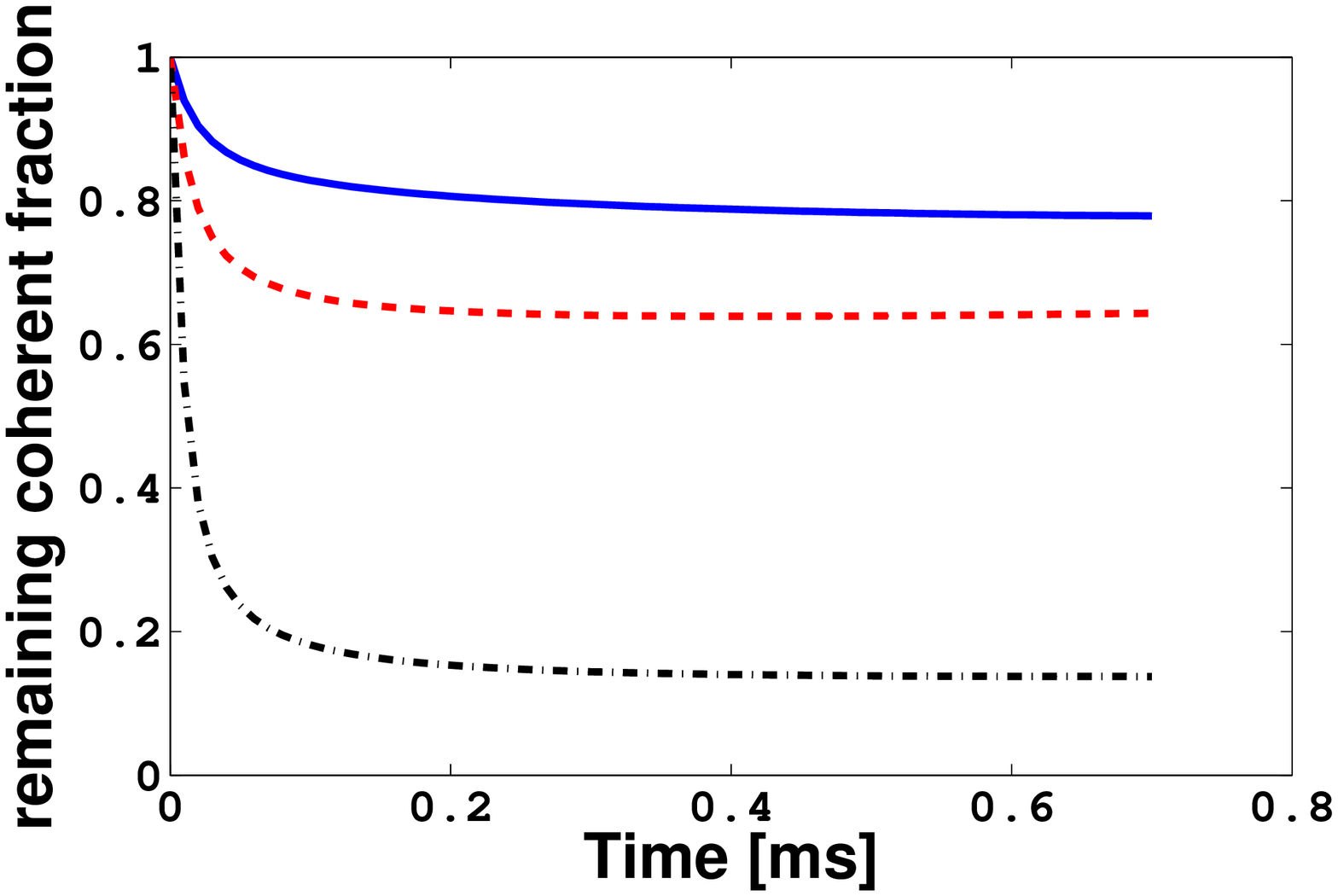}}
%\subfigure[]{
\caption{(a) Measured and calculated surviving fraction of coherent atoms, as a function of lattice depth. The averaged experimental data points and errorbars (depicting the error of the mean) are plotted in blue.
The solid blue line connecting the experimental points serves as a guide to the eye. The data clearly shows decay enhancement for deep lattices. The calculation is given by the dashed-red curve (see text for details). The dashed-dotted black curve depicts the GR result, showing no decay for $V_{lat} > 4 E_R$. Inset: Sequence of excitation and decay measurements which gives the surviving fraction of coherent atoms $f_c (t)$ (see text). Following this sequence the magnetic trap is switched off at $t=1000 \mu$sec, and a time-of-flight image is taken following $38$msec of free expansion. (b) Calculated surviving fraction of coherent atoms as a function of time, for 3 different lattice depths ($V_{lat}=5E_R$ - solid blue, $V_{lat}=10E_R$ - dashed red, $V_{lat}=20E_R$ - dashed-dotted black).} \label{fig:col_res}
\end{figure}

The decay of the excitation is measured using the pulse sequence depicted in the inset of Fig. \ref{fig:col_res}(a). The lattice beams are turned on and off adiabatically over $250 \mu$s, and kept fixed for $500 \mu$s. A Bragg excitation pulse at a frequency difference of $6$kHz and Rabi frequency of $1$kHz is applied during the first $100 \mu$s after the lattice turn-on, allowing for $400 \mu$s of decay \footnote{See Supporting Information, Sec. II, par. 4.}. As a normalization, the experiment is repeated with a Bragg pulse applied during the last $100 \mu$s of the lattice existence, such that no time is left for decay. 
The ratio of the number of atoms remaining excited in the first experiment ($P_f$) to that recorded in the second one ($P_i$), measured from time-of-flight images such as Fig. \ref{fig:setup}b \cite{katz_beliaev} (taken by switching the trap off immediately following the pulse sequence and allowing $38$ msec of expansion) gives the surviving fraction of coherently excited atoms $f_c (t)$:
\begin{equation}
f_c (t) = \frac{P_f (t)}{P_i} = \frac{P_i e^{-\int_0^t dt' R(t')}}{P_i} = e^{-\int_0^t dt' R(t')}. \label{f_c}
\end{equation}
Here $R(t)$ is the time-dependent rate of decay. This normalization method rejects ``common-mode'' noise in the experiment.
%The decay of the excitation is measured using the pulse sequence depicted in Fig. \ref{fig:col_res}(a). The measured population of excited atoms
%after decay, $P_f$, is generally dependent on the initial population created in the Bragg pulse $P_i$, which in turn varies with lattice depth.
%In order to correctly normalize the experimental results such that they do not depend on $P_i$, we measure for each lattice depth both $P_i$ and
%$P_f$ and find
%\begin{equation}
%f_c (t) = \frac{P_f (t)}{P_i} = \frac{P_i e^{-\int_0^t dt' R(t') t'}}{P_i} = e^{-\int_0^t dt' R(t') t'}. \label{f_c}
%\end{equation}
%Here $R(t)$ is the time-dependent rate of decay. This is the ratio between the number of atoms that have not decayed after being excited in the
%first $100 \mu$s following the lattice switch-on ($P_f$), to those excited in the last $100 \mu$s, before the lattice switch-off ($P_i$).
%This method rejects ``common-mode'' noise in the experiment, thus providing better signal-to-noise ratio in the measurement of decay.

Figure \ref{fig:col_res}(a) presents the measured fraction of coherently excited atoms that did not decay, as a function of lattice depth. It can be seen that the decay rate increases with the lattice depth. As shown below, this result completely violates the Golden-Rule (GR) behavior \cite{griffin}, but agrees with a decay formalism that includes the effects of the short-time non-exponential regime.

The decay rate $R$ of a state whose unperturbed energy is $E_0$ into a reservoir is usually calculated using the Fermi Golden Rule \cite{sakurai},
\begin{equation}
R_{GR} \left( E_0 \right) = \frac{2 \pi}{\hbar} \rho \left( E_0 \right) |V \left( E_0 \right) |^2, \label{fgr1}
\end{equation}
where $\rho(E)$ is the reservoir density of states and $V(E)$ is the system-reservoir coupling matrix element. The GR presumes the Markov limit of observation times much longer than the correlation (memory) time $t_c$ of the reservoir response. 
%By varying the lattice depth we change the excitation spectrum, altering $E_0$, $\rho(E)$ and $V(E)$, ultimately affecting the decay rate.

%For deep enough lattices, the lowest band shape becomes completely convex. This convexity has been predicted to preclude decay of excitations \cite{griffin} (as in other systems \cite{landau_maxon,hills_maxon}), a prediction which does not hold for our experiment (see below).
%It is important to note that this convexity is a signature of superfluids \cite{landau_maxon,hills_maxon}), whose dissipationless flow stems from the linearity of the spectrum at low momenta.

The change in the excitation spectrum as a function of lattice depth affects the decay rate in a manner that can be understood in terms of momentum and energy conservation of the decay process. The excited $q=0.9 k_L$ mode can decay into a pair of modes, such that the sum of their momenta and energy equals the momentum and energy of the excitation. A {\em concave} spectrum, as in weak lattices ($V_{lat} \simleq 4 E_R$, see Fig. \ref{fig:spect}), implies that the decay products have less energy than the excitation, and therefore decay is possible, as excess energy is carried by transverse excitations. By contrast, in deep lattices ($V_{lat} \simgeq 4 E_R$), the completely convex spectrum (Fig. \ref{fig:spect}) implies that the excitation has less energy than any pair of decay products \cite{griffin}. In this scenario the excitation can be described as a quasi-particle with dissipationless flow, analogous with the maxon quasi-particle of liquid He \cite{landau_maxon}. This effect completely {\em inhibits} the coupling between the excitation and the available modes for decay at {\em long times} (compatible with the GR - see black dashed-dotted line in Fig. \ref{fig:col_res}a).

However, for finite-time decay, for which the energy-conservation constraint is relaxed through time-energy uncertainty, the above GR-based results are no
longer valid. 
%In essence, for short enough times, the Golden-Rule decay rate does not hold, which is a manifestation of non-exponential decay. In our experiment finite-time effects are significant, since the excitation spectrum flattens out as the lattice depth increases.
Since the lowest band narrows down exponentially with lattice depth (Fig. \ref{fig:spect}), for any finite duration of an experiment there is a lattice depth beyond which the decay rate calculated by the GR (Eq. (\ref{fgr1})) strongly deviates from the actual result. 
%Eq. (\ref{fgr2}). This implies that the decay is non-exponential, and full account of the decoherence must allow for both the finite duration of the excitation, and the change in the reservoir spectral response as a function of the lattice depth. Non-exponential decay at short times occurs in all systems. However, in an excited atom spontaneously decaying to the ground-state, for example, this short-time limit is on the order of {\em atto-seconds} \cite{kurizki}, rendering deviations from exponential decay utterly negligible. 
In our system the reservoir correlation time $t_c$ is exceptionally long ({\em hundreds of microseconds}), so that the integrated decay probability over the experimental time duration drastically deviates from the GR exponential decay \cite{mazets}. The timescale of the reservoir correlation time is dictated by the bandwidth of the available modes for decay. In our system this bandwidth is given by the width of the lowest Bloch band, which is on the order of a few kHz.

The above account can be given in terms of a more general (``universal'') expression than Eq. (\ref{fgr1}) \cite{kurizki}. The universal expression allows for arbitrary observation times, including $t << t_c$, and hence the entire spectral dependence of the excitation and the reservoir response:
\begin{equation}
R_{E_0}(t) = 2 \pi \int_0^\infty F \left( \omega - E_0/\hbar,t \right) G(\omega) d \omega. \label{fgr2}
\end{equation}
This expression relates the decay rate to the overlap (convolution) of two spectral functions: $F \left( \omega - E_0/\hbar,t \right)$, the time-dependent spectral function of the excitation centered at the unperturbed energy, $E_0$, and $G (\omega)$, the response spectrum of the reservoir, which is the Fourier transform of its non-Markov memory functional $\Phi(t)$ that dies out at $t \geq t_c$. Control of the decay is achievable by {\em altering} this overlap.
Here we explore this simple expression for an interacting many-body system, as compared to the non-interacting atoms studied before \cite{raizen}. We combine reservoir engineering of $G(\omega)$ with system probing at short times $t \leq t_c$ that affects $F \left( \omega - E/\hbar \right)$, so as to achieve significant changes in the overlap of the two functions and therefore of the decay.
% (see also \cite{ketterle_zeno,sakurai,itano,kwiat}).
%This is the first demonstration of such effects as concerns the control of decay of many-body excitations.
%This universal formula captures the physics behind the quantum Zeno and anti Zeno effects\cite{kurizki}, as well as that of reservoir engineering. In essence, Eq. (\ref{fgr2}) relates the decay rate to the overlap of the two spectral functions of the system excitation or modulation and of the reservoir response. Control of decay is achieved by altering the overlap between the two functions. Shifting the two spectral functions to the point of no overlap effectively decouples the system from the continuum, thus suppressing decay. Conversely, such shifting may increase the overlap, and hence the decay rate.

The reservoir spectral response function can be written as:
\begin{equation}
G (\omega_{{\bm k},{\bm q}}) = \rho (\omega_{{\bm k},{\bm q}}) |A_{{\bm k},{\bm q}}|^2, \label{Gw}
\end{equation}
where $A_{{\bm k},{\bm q}}$ are the Beliaev matrix elements given in terms of the Bogoliubov amplitudes $u_{jq}$ and $v_{jq}$ from Eq. (\ref{bloch_bogo}) \cite{katz_beliaev,ozeri}, and $\rho(\omega_{{\bm k},{\bm q}})=\left[ \partial \omega_{k,q} / \partial{(k,q)} \right]^{-1}$ is the density of states available for decay calculated from Eq. (\ref{bloch_bogo}) \cite{kramer}. As the lattice deepens, the Beliaev matrix elements increase, and the spectral response of the reservoir is significantly enhanced. However, in the infinite-time limit associated with the GR (Eq. (\ref{fgr1})),
the system's spectral excitation function is 
\begin{equation}
F(E_k/\hbar-\omega_{{\bm k},{\bm q}},t \rightarrow \infty) = \delta (E_k/\hbar-\omega_{{\bm k},{\bm q}}), \label{Fgr}
\end{equation}
i.e. it becomes localized at $E_k/\hbar$, the resonant frequency of the excitation. As the lattice depth increases, this excitation peak is pushed to lower energies, such that the spectral overlap between Eqs. (\ref{Gw}) and (\ref{Fgr}) goes to zero. Therefore, at the GR limit the decay rate of the excitation is expected to be completely
suppressed above a certain lattice depth \cite{griffin}.
By contrast, in our experiment, the spectral excitation function of the system is spread around $E_k/\hbar$
\begin{equation}
F (E_{\bm k}/\hbar - \omega_{{\bm k},{\bm q}},\Delta t) = \frac{sin^2[(E_{\bm k}/\hbar - \omega_{{\bm k},{\bm q}}) \Delta t]}{(E_{\bm k}/\hbar - \omega_{{\bm k},{\bm q}})^2 \Delta t}, \label{Fw}
\end{equation}
with a width ($1/\Delta t$) given by the inverse duration of the experiment.
The GR result fails if this width becomes comparable to the energy spread of the modes available for decay \cite{kurizki}, as in our experiment.

In Fig. \ref{fig:col_res}(a) we show the theoretical prediction (dashed-red line) for the surviving coherently-excited fraction of atoms $f_c$, as a function of lattice depth, alongside the experimental results. This is based on numerical integration of Eq. (\ref{f_c}) using Eqs. (\ref{fgr2}), (\ref{Gw}) and (\ref{Fw}), over the experimental time interval ($400 \mu$s). The theoretical curve agrees with the experimental results and stands in contrast to the GR limit, calculated by integrating Eq. (\ref{f_c}) over $400 \mu$s with $R(t) = R_{GR}$ (dashed-dotted black), which predicts that there is strictly no decay for the lattice depths of $V_{lat} \simgeq 4 E_R$ in Fig. \ref{fig:col_res}(a).
% (in agreement with \cite{griffin}).
Both theory and experiment indicate that, as the lowest band of the spectrum flattens out with increasing lattice depth, {\em finite-time effects dominate}
to cause substantial enhancement of decay, as compared to the infinite-time (GR) decay. This is the central result of this paper, indicating that Eq. (\ref{fgr2}) applies to our experiment. 

In Fig. \ref{fig:col_res}(b) we plot the calculated time dependence of the decay for different lattice depths. For all of these lattice depths GR theory predicts strictly no decay. These results show that the main part of the decay occurs in the first $100 \mu$s, which are difficult to access experimentally. We have also measured the remaining coherent fraction as a function of lattice depth for decay over $100 \mu$s and $250 \mu$s, and have found results similar to Fig. \ref{fig:col_res}(a), which is in accordance with the prediction of Fig. \ref{fig:col_res}(b).

The discrepancy between the experimental data and the theoretical curve for weak lattices could be partly attributed to the small background of thermal atoms. These thermal atoms act as a reservoir as well, and cause decay through the Landau damping process (see, e.g. [9]). This results in somewhat more decay than theoretically expected. For deep lattices, many-body effects cause depletion of the condensate on the order of a few percent, which could result in similar deviations from Bogoliubov theory. In addition, small non-adiabatic effects in this regime contribute to the discrepancy between the actual experiment and the theoretical prediction.
%In addition, the strong decay in this regime reduces the signal-to-noise ratio in the absorption imaging technique used in the experiment.
%The slight discrepancy between the experimental data and the theoretical curve could be partly attributed to the small background of thermal atoms.
%We note slight deviations of the experimental data from the theoretical prediction, which could be related to the existence of a background of low-temperature thermal atoms.

%The reason for enhanced decay can be traced to the overlap of the spectral functions of Eq. (\ref{fgr2}).
%If, as is the case here, $\left. \frac{dG}{d \omega_{k,q}} \right| _{\omega_{k,q}=E_k/\hbar} \neq 0$, i.e. the peak of $G(\omega_{k,q})$ does not coincide with the resonant energy $E_k$, then instead of QZE we have the AZE conditions \cite{kurizki}: the decay rate (\ref{fgr2}), expressed as the overlap of (\ref{Fw}) and (\ref{Gw}) increases as $\Delta t$ decreases. Generally speaking, $F \left( \omega - E/\hbar,t \right)$ can be broadened by measuring the system energy at intervals $t \leq t_c$ \cite{kurizki}. If the overlap is thereby diminished, typically at $t << t_c$, the decay rate is suppressed in QZE-like fashion. Conversely, if the overlap is enhanced, typically at $t \sim t_c$, the decay rate grows in AZE-like fashion\cite{kurizki,raizen}.

%Only for $\Delta t << t_c$, inaccessible for us, can QZE prevail.

This is the first observation of finite-time (AZE-like) decay enhancement of a coherent, {\em many-body} excitation decaying into its natural environment. QZE is much harder to observe since (as explained above) it may only occur in weak lattices (where the GR decay is nonzero) and there decay dynamics is hard to access. The results presented here attest to the universality of Eq. (\ref{fgr2}) and its adequacy for multipartite systems.
%Allegedly, decoherence precludes macroscopic coherence, since otherwise we would be required to {\em keep track of the system correlations with each individual degree of freedom of the environment}, a tantalizing and utterly impractical task.
We have demonstrated that {\em very limited knowledge} of the environment response (its correlation time $t_c$), suffices for controllable modification of decay in complex macroscopic systems, if
% that are embedded in noisy environments.
this control is faster than $t_c$.
Finally, this suggests the possibility of studying the quantum-classical transition and decoherence dynamics using long-lived coherent excitations, and may be extended to other systems and their environments.
%in various systems: double-site Bose-Hubbard systems or excitonic condensates, to mention but a few.

We acknowledge the support of ISF, GIF, DIP and EC (MIDAS STREP, FET Open).

%\bibliography{bibBEC3}

%% Here is the endmatter stuff: Supplementary Info, etc.
%% Use \item's to separate, default label is "Acknowledgements"

\end{document}